\documentclass{PoS}
\usepackage{subfigure}
\usepackage{xcolor}
\usepackage{amsmath}
\usepackage{mathptmx}
\usepackage{slashed}
\usepackage{cite}

\allowdisplaybreaks[1]
\newcommand{\GeV}{\,\mathrm{GeV}}
\newcommand{\MeV}{\,\mathrm{MeV}}

\newcommand{\Lag}{\mathcal{L}}
\newcommand{\LSM}{\mathrm{L}\sigma\mathrm{M}}
\newcommand{\LLSM}{\mathcal{L}_{\mathrm{L}\sigma\mathrm{M}}}
\newcommand{\wq}{\widetilde{\sigma}}

\newcommand{\ie}{\textit{i.e.\,}}
\newcommand{\eg}{\textit{e.g.\,}}
\newcommand{\cf}{\textit{cf.\,}}
\newcommand{\Ord}[1]{\mathcal{O}(#1)}
\newcommand{\Tc}{T_{\text{CP}}}
\newcommand{\Tpc}{T_{\text{pc}}}
\newcommand{\muc}{\mu_{\text{CP}}}

\title{Photon emission rates near the critical point in the linear sigma model}

\ShortTitle{Photon emission rates near the critical point in the linear sigma model}

\author{F. Wunderlich	\\
        Helmholtz-Zentrum Dresden-Rossendorf, Institute of Radiation Physics, 
        \mbox{Bautzner Landstr. 400}, D-01328 Dresden, Germany\\
        and \\
        Institut f\"ur Theoretische Physik, Technsche Universit\"at Dresden, D-01062 Dresden, Germany\\
        E-mail: \email{f.wunderlich@hzdr.de}}

\author{\speaker{B. K\"ampfer}\\
        Helmholtz-Zentrum Dresden-Rossendorf, Institute of Radiation Physics, 
        \mbox{Bautzner Landstr. 400}, D-01328 Dresden, Germany\\
        and \\
        Institut f\"ur Theoretische Physik, Technsche Universit\"at Dresden, D-01062 Dresden, Germany\\
        E-mail: \email{kaempfer@hzdr.de}}
        
\abstract{Employing the linear sigma model, the effective masses of quasi-particle excitations are found
         to exhibit significant variations within the phase diagram, which has a critical point at non-zero
         chemical potential, where a first-order phase transition sets in.
         Soft-photon emission rates in lowest order display, for selected channels, a sensible dependence
         on the effective masses of the involved excitations and let us argue that they could map out the 
         phase diagram.}

\FullConference{9th International Workshop on Critical Point and Onset of Deconfinement - CPOD2014,\\
		17-21 November 2014\\
		ZiF (Center of Interdisciplinary Research), University of Bielefeld, Germany}

\begin{document}
\section{Introduction}
Confinement and spontaneous symmetry breaking as well as their respective counterparts - deconfinement and chiral 
restoration - are central issues of QCD. Their mutual relationship is among the main questions investigated in 
heavy ion collisions (HICs). A particularly challenging task is the establishment of the QCD phase diagram in a 
region of temperature $T$ and baryo-chemical potential $\mu$ which is accessible in HICs. 
While the beam energy scan at RHIC and forthcoming accelerator/detector installations, \eg at FAIR/CBM and NICA/MPD,
aim at finding signatures of the onset of deconfinement as a first-order phase transition in a critical point (CP),
QCD based \cite{Karsch:2003va,Fodor:2004nz} and QCD rooted \cite{Fischer:2014vxa} calculations try to quantify 
its coordinates.
Instead of facing the sign problem of QCD at non-zero baryon density, one can resort to suitable models which display
a CP at $(\Tc, \muc)> (0,0)$ and a line of first-order phase transitions for $\mu>\muc$ 
(\cf \cite{Fukushima:2010bq,Fukushima:2013rx} for recent surveys). By universality arguments,
one can elucidate observables signaling the CP. 

In the previous search for deconfinement effects in HICs,
electromagnetic probes (\cf \cite{Vujanovic:2013jpa,Shen:2013vja,Lee:2014pwa} for recent evaluations and further references)
have been considered as promising since, by their penetrating nature, they monitor the 
full space-time evolution of matter in the course of HICs.
Given the proximity of deconfinement and chiral restoration at $\mu/T\ll 1$, the dilepton spectra have been considered 
in \cite{Rapp:1999ej,Rapp:2009yu} as candidates for messengers of chiral restoration. It appeared that at the pseudo-critical
temperature of $\Tpc\sim 154\MeV$ the dilepton emission rates of confined matter (hadrons) and deconfined matter 
(quarks and gluons) become degenerate - quite natural since $\Tpc$ quantifies the cross over location at $\mu/T\ll 1$.
This degeneracy can be named quark-hadron duality and has been employed \cite{Rapp:1999zw,Gallmeister:1999dj,Gallmeister:2000ra} to verify a 
``thermal radiation'' component besides 
hard initial yield and electromagnetic hadron final-state decays after the disassembly of the fireball in HICs.
The approximate degeneracy of real-photon emission rates of confined and deconfined matter has been pointed out quite 
early \cite {Kapusta:1991qp} and can be used analogously to arrive  at a consistent modeling of thermal
radiation in HICs, either
schematically \cite{Gallmeister:2000si,Thomas:2005wca} or with many refinements \cite{Rapp:2014qha,vanHees:2014ida,Heffernan:2014mla}.

The recent analysis in \cite{Rapp:2014qha,Gale:2014dfa}, however, revealed that the calculated photon spectra fall short in comparison
with data. The tension is caused, to some extent, by the consistency requirement of photon-$v_2$ and photon-$p_\bot$ 
systematics. According to \cite{Rapp:2014qha,vanHees:2014ida}, a solution is offered by the hypothesis of a 
``pseudo-critical enhancement'' of the photon
emission at $\Tpc\pm\Delta T$ with $\Delta T = \Ord{15\MeV}$. It is therefore tempting to look for arguments
which could support the hypothesis launched there and, more general, to extend the consideration over the 
QCD phase diagram and to check whether the CP can have a distinguished impact on the photon emission rate.

\section{Linear $\sigma$ model}
\begin{figure}
   \centering
   \subfigure{\includegraphics[height = 119pt,clip=true,trim=2mm 2mm 19mm 16mm]
             {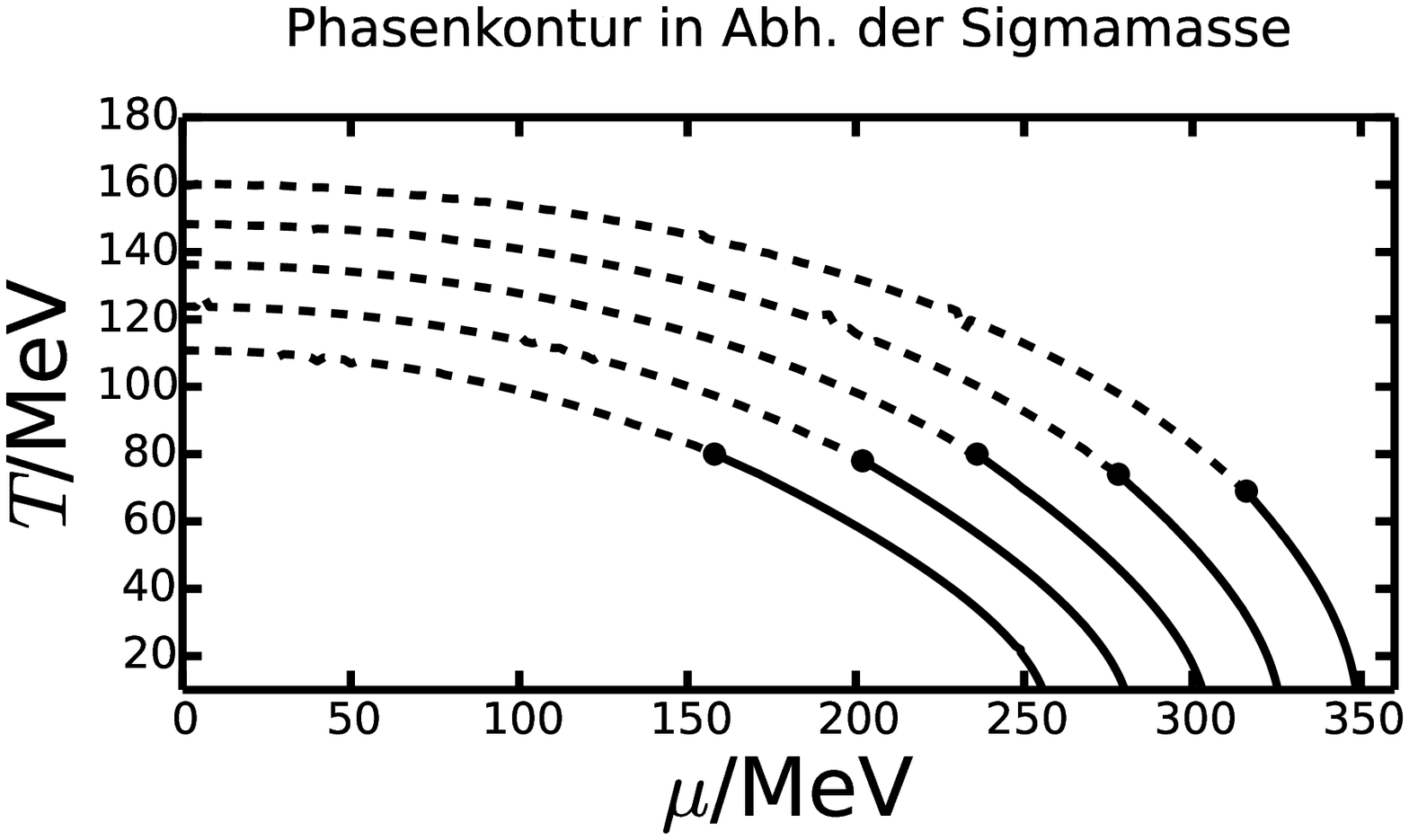}
%             \put(-23, 100){\fcolorbox{white}{white}{(a)}}
             }
   \subfigure{\includegraphics[height = 119pt,clip=true,trim=25mm 2mm 19mm 16mm]
             {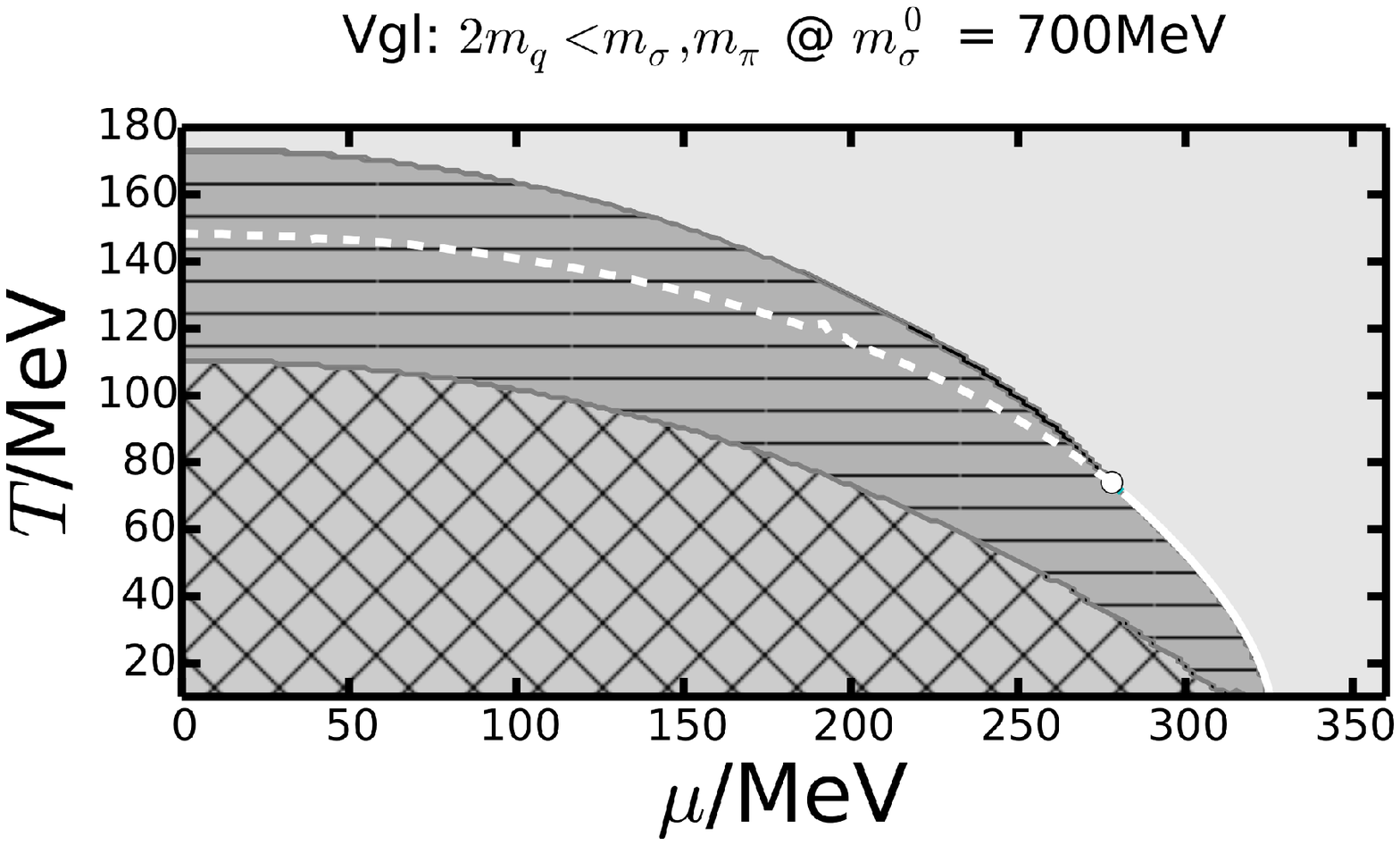}
 %            \put(-23, 100){\fcolorbox{black}{white}{(b)}}
             \setlength{\fboxsep}{1.5pt}
             \put(-191, 32){\colorbox{white}{\footnotesize{$m_\sigma > 2m_\psi > m_\pi$}}}
             \put(-191, 83){\colorbox{white}{\footnotesize{$2m_\psi > m_{\pi,\sigma}$ }}}
             \put(-7, 100){\makebox[0pt][r]{\footnotesize{$2m_\psi < m_{\pi,\sigma}$}}}
             \put(-7, 82) {\makebox[0pt][r]{\footnotesize{$m_\pi > 2m_\psi > m_\sigma$}}}
             \put(-37, 78){\thicklines\vector(-2,-1){20}}
             \label{fig:massenvgl_mesonen_quark}}
   \caption{Left panel: Phase structure of the $\LSM$ for various values of the sigma mass parameter 
   $m_\sigma^\text{vac} =$ 400, 500, 600, 700 and 800\,MeV (bottom to top). The solid curves
   depict first-order phase transition lines and the dotted curves are an estimate for a pseudo-critical temperature 
   $\Tpc(\mu)$. Both curves meet at the CP (dots).
   Right panel: Relations of the effective masses $m_{\psi, \pi, \sigma}$ over the phase plane 
   (for $m_\sigma^\text{vac} = 700\MeV$, with phase structure as in left panel).
   }
   \label{fig:m_si_scan}
   \label{fig:massenvgl}
\end{figure}

We utilize here the linear sigma model ($\LSM$), which has been originally designed to model spontaneous chiral symmetry
breaking \cite{GellMann:1960np} and has been later coupled to the fermionic sector. 
It is a special quark-meson model with the Lagrangian
$  \LLSM = \Lag_{\partial \psi} + \Lag_{\psi M} + \Lag_{\partial M} + \Lag_M + \Lag_{s.b.}$, 
where the kinetic terms $\Lag_{\partial \psi}$ and $\Lag_{\partial M}$ are supplemented by the $O(4)$ symmetric meson
self-interaction $\Lag_{M}$, which refers to an $SU(2)_L\times SU(2)_R$ symmetry which, in combination with the symmetry 
breaker $\Lag_{s.b.}=H\sigma$, dynamically generates mass terms; the fermion-meson coupling of Yukawa type is encoded in 
$\Lag_{\psi M}$.
The field content of $\Lag$ is a two component Fermion (light quark doublet $\psi$), an iso-vector ($\vec\pi$) and an 
iso-scalar ($\sigma$) meson. 
(The model can be extended to SU(3) and may include vector mesons \cite{Beisitzer:2014kea} in the spirit of the 
gauged $\LSM$ \cite{Gasiorowicz:1969kn} or can be supplemented by some gluon dynamics encoded in the Polyakov loop 
\cite{Schaefer:2007pw} or a glueball condensate \cite{Sasaki:2011sd}.)
When including linearized meson field fluctuations, as in \cite{Mocsy:2004ab,Bowman:2008kc,Ferroni:2010ct},
the resulting phase diagram has, in fact, a line
of first-order phase transitions (\cf solid curves in Fig.~\ref{fig:m_si_scan}), where the 
pressure as a function of $T$ and $\mu$ has a self-intersection defining $T_c(\mu)$. 
The fine structure of the region, where the CP is located \cite{Ferroni:2010ct}, is thought to reflect the restriction to 
linearized fluctuations. (For a better account of fluctuations within a FRG approach, \cf \cite{Tripolt:2013jra}.)

The parameters in the Lagrangian are commonly fixed by requiring a coincidence of the dynamically generated  meson 
quasi-particle masses in the vacuum ($T=\mu=0$) with the experimental
values of pion, $m_\pi^\text{vac} = 138\MeV$, and sigma, $m_\sigma^\text{vac} = 700\MeV$, and the constituent quark mass,
adjusted to one third of the 
nucleon mass, $m_\psi^\text{vac} = 312\MeV$. The remaining free parameter is determined by 
$\langle \sigma \rangle_\text{vac} = f_\pi$.
With these settings, the CP coordinates are $(\Tc, \muc) \sim (74\MeV,278\MeV)$ (without the Fermion vacuum loop; for its
impact \cf \cite{Skokov:2010sf}).
Since the sigma field in $\LLSM$ can be considered as an effective  field which may be tentatively identified with 
$f_0(600)$ leading to vacuum mass values of $m_\sigma^\text{vac} = 400 - 800\MeV$. It is instructive to 
study its impact on the phase diagram (\cf~left panel of Fig.~\ref{fig:m_si_scan}). 
It turns out that the phase contour curves are shifted approximately linearly with $m_\sigma^\text{vac}$. 
For an estimate we note $\Delta \Tpc \approx 0.12 \Delta m_\sigma$ on the temperature axis. The CP is
shifted approximately parallel to the $\mu$ axis by about $40\MeV$ per $100\MeV$ change of $m_\sigma^\text{vac}$ in this 
mass region. The impact of $m_\pi^\text{vac}$ is discussed extensively in \cite{Bowman:2008kc}. We refrain here from a 
discussion of the phase structure under variations of all parameters at fixed $\Tpc$, which we define here as the curve,
where the normalized heat capacity as a function of $T$ at fixed $\mu$ has a maximum - other definitions of the 
pseudo-critical temperature are conceivable. For further discussion of the phase diagram within effective chiral
quark-meson models \cf \cite{Schaefer:2006ds,Nakano:2009ps,Schaefer:2011ex}.

\begin{figure}
   \centering
   \subfigure{{\includegraphics[height = 120pt,clip=true,trim=5mm 8mm 10mm 24mm]
             {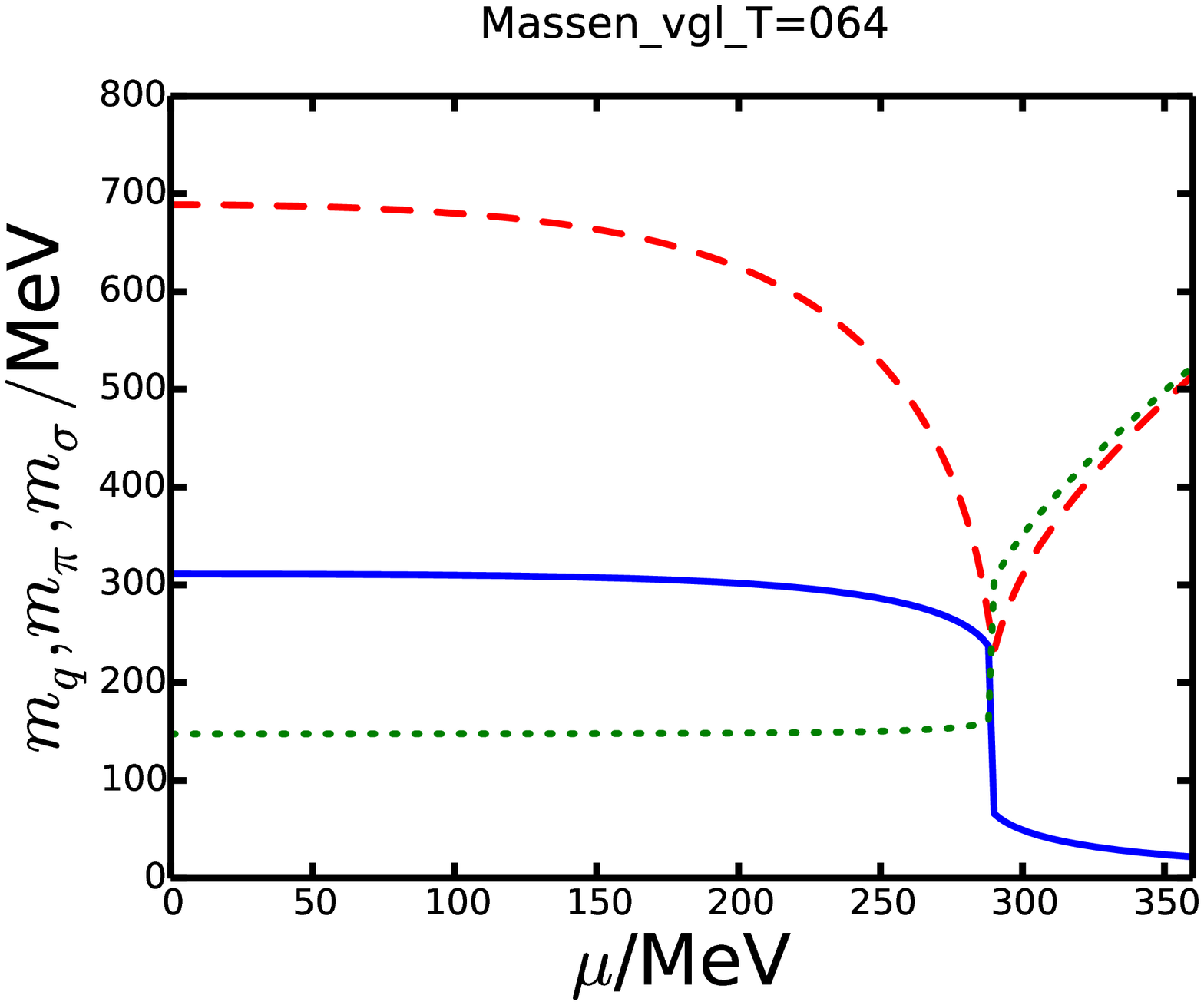}}
             \put(-127, 65){\scriptsize $T=\Tc-10\MeV$}
             \label{masses_T=64}}
   \subfigure{{\includegraphics[height = 120pt,clip=true,trim=30.5mm 8mm 10mm 24mm]
             {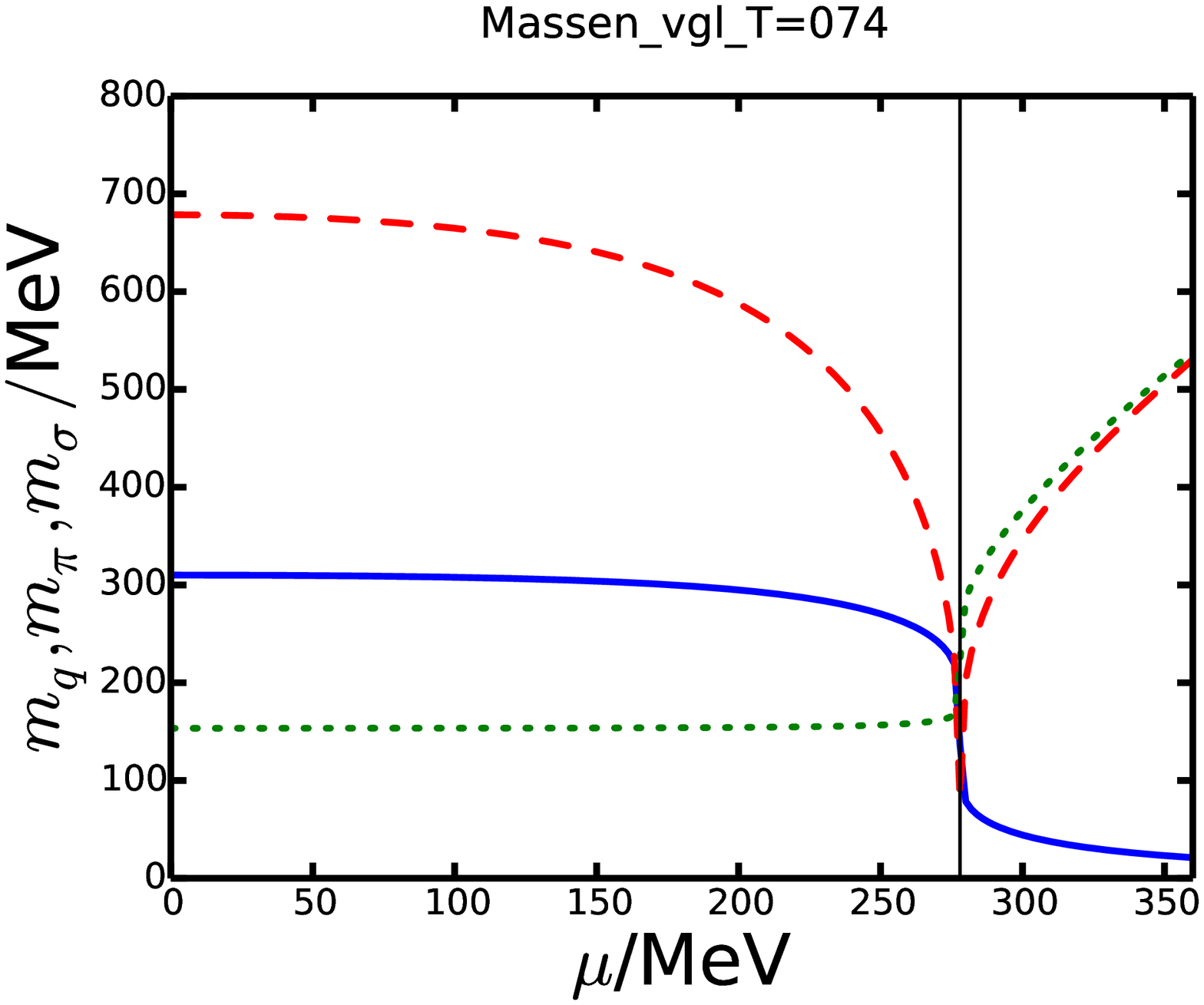}}
             \put(-127, 65){\scriptsize $T=\Tc$}
             \label{masses_T=74}}
   \subfigure{{\includegraphics[height = 120pt,clip=true,trim=30.5mm 8mm 10mm 24mm]
             {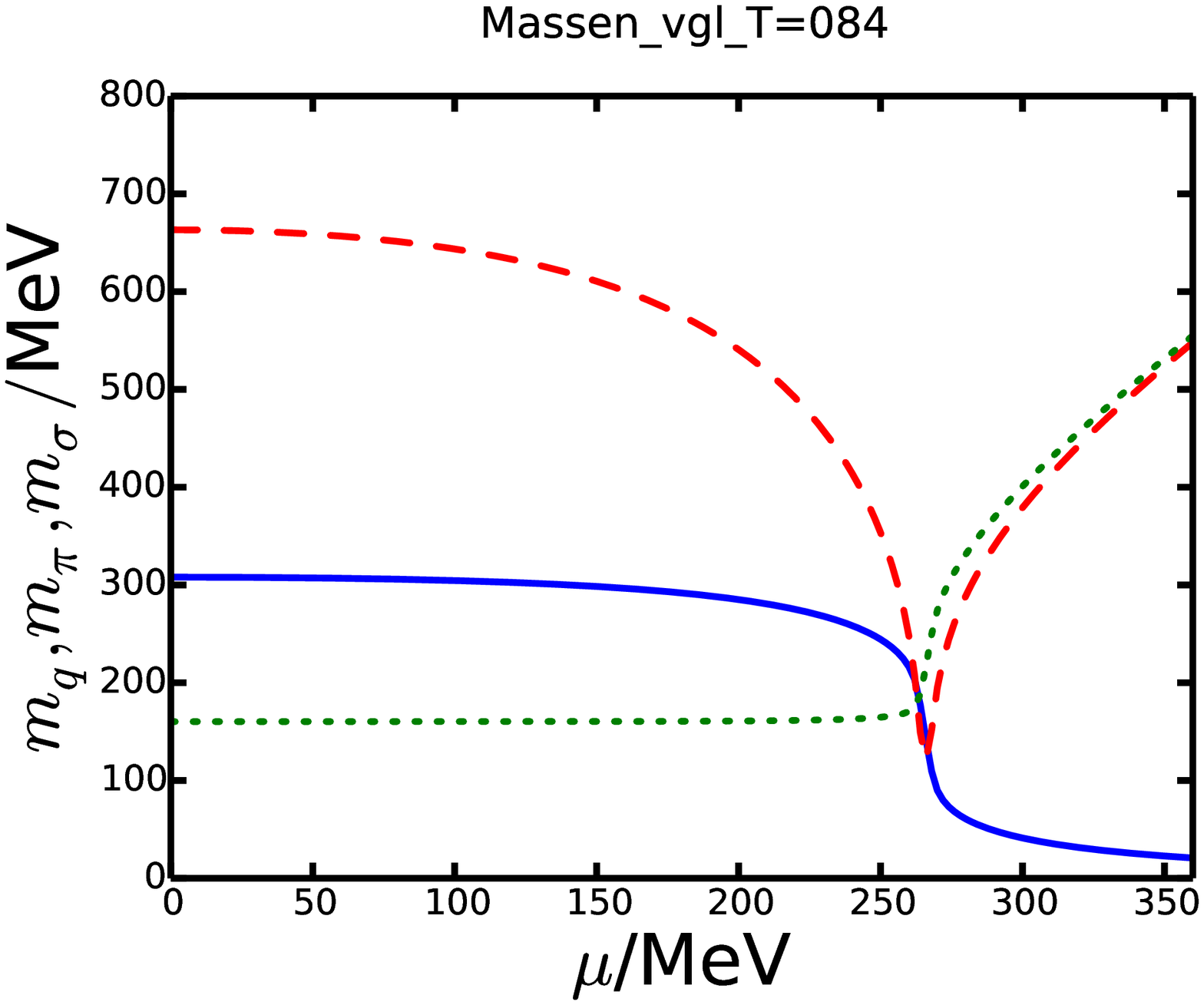}}
             \put(-127, 65){\scriptsize $T=\Tc+10\MeV$}
             \label{masses_T=84}}
   \caption{Quark (blue solid curve), sigma (red dashed curve) and pion (green dotted curve) effective masses as 
            a function of the quark chemical potential $\mu$ for 
            temperatures slightly below (left panel), at (middle panel) and slightly above (right panel)
            the critical temperature $\Tc$. The position of the critical chemical potential $\muc$ is depicted
            by the thin vertical line in the middle panel.}
   \label{fig-massen}
\end{figure}

Analogously to the mean field approximation \cite{Scavenius:2000qd} a crucial role is played by dynamically generated mass 
terms, which may be considered
as quasi-particle masses $m_{\psi,\pi,\sigma}$ (\cf Fig.~\ref{fig-massen}) of thermal excitations of the
$\psi$, $\vec\pi$ and $\sigma$ fields (for details see 
\cite{Wunderlich:2014cia}). Prominent features are (i) degeneracy of $m_\pi$ and $m_\sigma$ at temperatures larger than $\Tpc(\mu)$
or $T_c(\mu)$, respectively, (ii) a rapid dropping of $m_\psi$ at $T=T_c(\mu)$ and a mild dropping at 
$T\sim \Tpc(\mu)$ and (iii) a global minimum of $m_\sigma$ at $(\Tc, \muc)$ (for the rationale of the softening of the 
$\sigma$-mode, \textit{cf., e.g.},\cite{Yokokawa:2002pw,Fukushima:2002mp,Kitazawa:2014sga} and for the impact on di-photon spectra
\cite{Rehberg:1997xe,Volkov:1997dx}). 

It is remarkable that the mass ordering changes over the $T$-$\mu$ plane, see Fig.~\ref{fig:massenvgl} (right panel). The 
delineation of $2m_q > m_{\pi,\sigma}$ vs. $2m_q < m_{\pi, \sigma}$ coincides with $T_c(\mu)$. There is a narrow valley,
wherein $m_\sigma < 2m_q < m_\pi$ separating $2m_q>m_{\pi,\sigma}$ vs. $2m_q < m_{\pi,\sigma}$ regions. This and the other
delineation curves have no direct relation to $T_c(\mu)$ or $\Tpc(\mu)$.

\section{Photon emission rates}
In a phenomenological approach and as a first exploratory step one can consider the above mentioned quasi-particle masses 
as masses of the quanta entering the respective dispersion relations 
$E^2_{\psi,\pi,\sigma} = {\vec p}^2 + m^2_{\psi,\pi,\sigma}$, which in turn show up in the thermal occupation functions 
(\cf appendix B in \cite{Tripolt:2013jra}).
Due to the strong variations of these masses over the phase diagram (\cf Fig.~\ref{fig-massen}), one can expect also a 
non-trivial pattern of the real-photon emission rate (see figure 2 in \cite{Wunderlich:2014cia}). 

Leaving a theoretically sound approach from Wightman functions to imaginary part of the retarded photon propagator to 
kinetic 
theory expressions from cutting rules for separate work, we employ the phenomenologically anticipated kinetic approach by 
estimating the rates via
\begin{eqnarray}
   \omega \frac{d^7N}{d^4xdk^3} = C\int  \frac{d^3p_1}{2 p^0_1}\frac{d^3p_2}{2 p^0_2}\frac{d^3p_3}{2 p^0_3} 
                              f_1(p_1)f_2(p_2)(1\pm f_3(p_3))|M_{12\rightarrow3\gamma}|^2\delta^{(4)}(p_1+p_2-p_3-k),
                              \quad\label{rate-01}
\end{eqnarray}
with $C =(2(2\pi)^8)^{-1}$,
\ie we account for lowest-order tree-level $2\rightarrow 2$ processes $1+2\rightarrow 3 + \gamma$, where the photons
are minimally coupled to $\LLSM$ according to \cite{Mizher:2010zb}. 
These are annihilations $\psi_i + \bar \psi_j\rightarrow M + \gamma$ and Compton processes 
$\psi_i + M \rightarrow \psi_j + \gamma$ 
and $\bar \psi_i + M \rightarrow \bar \psi_j + \gamma$, respectively, with 
$i,j\in\{ u,d\}$ and $M\in\{\pi^0,\pi^\pm,\sigma\}$ being the respective meson(s) fulfilling charge conservation depending 
on the choice of $i,j$. 
(Clearly, this list of reactions is by far not exhaustive when having in mind hadronic sources in general. For instance, in 
\cite{Turbide:2003si} the non-linear $\sigma$ 
model with vector and axial-vector mesons has been used to investigate real-photon emission rates from hadron sources.
In \cite{Liu:2007zzw}, the channel $\pi^+\pi^-\rightarrow \sigma/\rho \rightarrow \pi^+\pi^-\gamma$ has been identified 
as important for soft photons. This and many more channels should be considered when attempting a comprehensive rate 
estimate.)
The matrix elements for Compton and annihilation processes are related by crossing symmetries.
The evaluation of \eqref{rate-01} by means of the $\delta^{(4)}$ and exploiting the cylinder symmetry in the center of 
mass frame of the reaction leaves four integrals
\cite{Alam:1999sc,Yaresko:2010xe} to be done numerically.

\section{Photon spectra}
An instructive form of \eqref{rate-01} can be derived by resorting to the Boltzmann approximation:
\begin{eqnarray}
   \omega \frac{d^3N}{dk^3} &\approx& \widetilde{C} e^{\epsilon \mu/T}\int_{s_0}^\infty ds \frac{\wq(s)}{s-m_3^2} e^{-F(s)/T}\label{Boltzmann_01}
\end{eqnarray}
with $\widetilde{C} = T/(32\omega (2\pi)^6)$, $\wq(s) = \int^{t_+}_{t_-} dt |M_{12\rightarrow3\gamma}|^2$, 
$t_\pm=m_1^2-\frac{s-m_3^2}{2s}\Big((s+m_1^2-m_2^2)\mp\sqrt{\lambda(s, m_1^2, m_2^2)}\Big)$,
$\lambda(x,y,z) = x^2 + y^2 + z^2 - 2xy -2xz -2yz$, $F(s) = (s-m_3^2)/(4\omega) + (s\omega)/(s-m_3^2) $
and $\sqrt{s_0} = \max\{m_1 + m_2, m_3\}$. The variable $\epsilon$ discriminates between three reaction types
under consideration: annihilation (\mbox{$\epsilon=0$}), Compton scattering at quarks (\mbox{$\epsilon = +1$}) 
and Compton scattering at antiquarks (\mbox{$\epsilon = -1$}).
For disentangling effects of the phase space distributions $f_{1-3}$, in particular the role of the involved quasi-particle masses,
and matrix elements $M_{12\rightarrow3\gamma}$ encoded in $\wq(s)$ let us analyze the function $F$.
Two cases are to be distinguished: (i) If the position of the minimum of $F$ w.r.t.\,$s$,
$s_1 = m_3^2 + 2\omega m_3$,  lies within the range of integration, \ie $s_0<s_1$, a Taylor expansion 
$F(s) = F(s_1) + 1/2 \cdot F''(s_1) \cdot (s-s_1)^2 + \Ord{(s-s_1)^3}$ is in order and the exponential in \eqref{Boltzmann_01}
can be replaced by a Gaussian with width $\sqrt{4 m_3 \omega^2 T}$ and height $\exp\{-(\omega+m_3)/T\}$,
making the exponential thermal suppression apparent. 
(ii) If $s_1 < s_0$, a Taylor expansion $F(s) = F(s_0) + F'(s_0) \cdot (s-s_0) + \Ord{(s-s_0)^2}$ yields a $s$ independent
suppression factor $\exp\{-F(s_0)/T\}$ that develops a maximum w.r.t. $\omega$ at
$\omega_p = (s_0-m_3^2)/(2\sqrt{s_0})$, meaning the spectrum displays a peak at $\omega_p>0$. Whether it appears is
thus controlled by the relative size of $s_0$ and $s_1$. If $m_1+m_2<m_3$, then $s_0=m_3^2<s_1$ (\cf case (i)), 
and the photon rate has its maximum at $\omega=0$ and declines exponentially with
power law corrections originating from $\wq(s)$ and other omitted factors in \eqref{Boltzmann_01}.
On the other hand, if $m_1+m_2>m_3$, then $s_1<s_0$ (\cf case (ii)), and the rate has a maximum 
$\propto e^{-(m_1 + m_2)/T}$ at $\omega_p$. Furthermore, even for $m_1+m_2>m_3$, there will
be a certain photon frequency $\omega_s>\omega_p$ for which $s_0=s_1$, implying that for $\omega>\omega_s$ case (i) is 
applicable and the rate behaves as $\propto \exp\{-\omega/T\}$ at large $\omega$.
Thus the mass ordering determines some basic features of the spectra, which is the reason why we have shown in 
the right panel of Fig.~\ref{fig:massenvgl} the corresponding regions.

\begin{figure}[htp]
   \centering
   \subfigure{{\includegraphics[width = 0.49\textwidth,clip=true,trim=4mm 2mm 4mm 16mm]
             {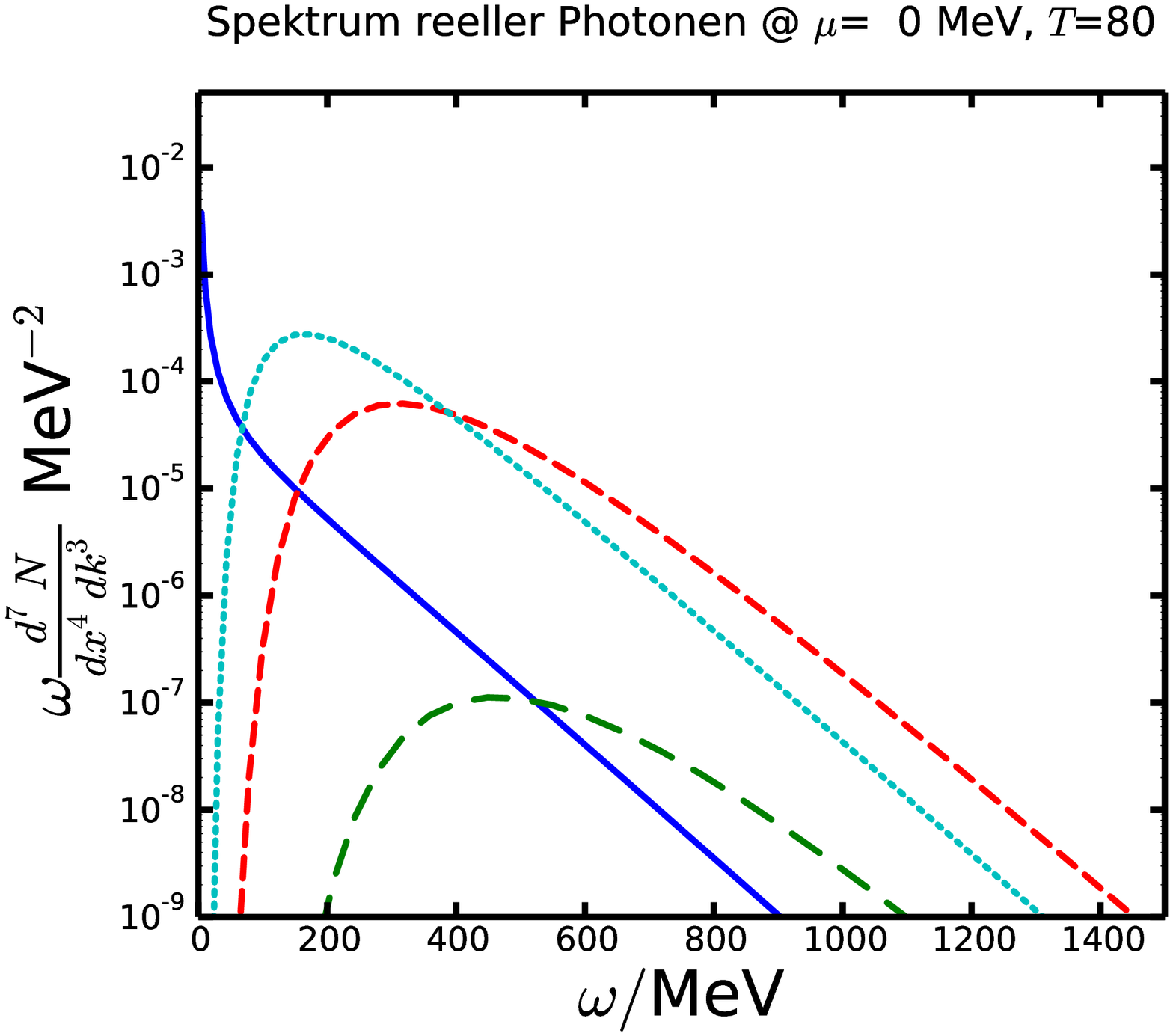}}
             \put(-93, 158){(a)}
             \label{subfig-Spektrum-000-80}}\hfill
   \subfigure{{\includegraphics[width = 0.49\textwidth,clip=true,trim=4mm 2mm 4mm 16mm]
             {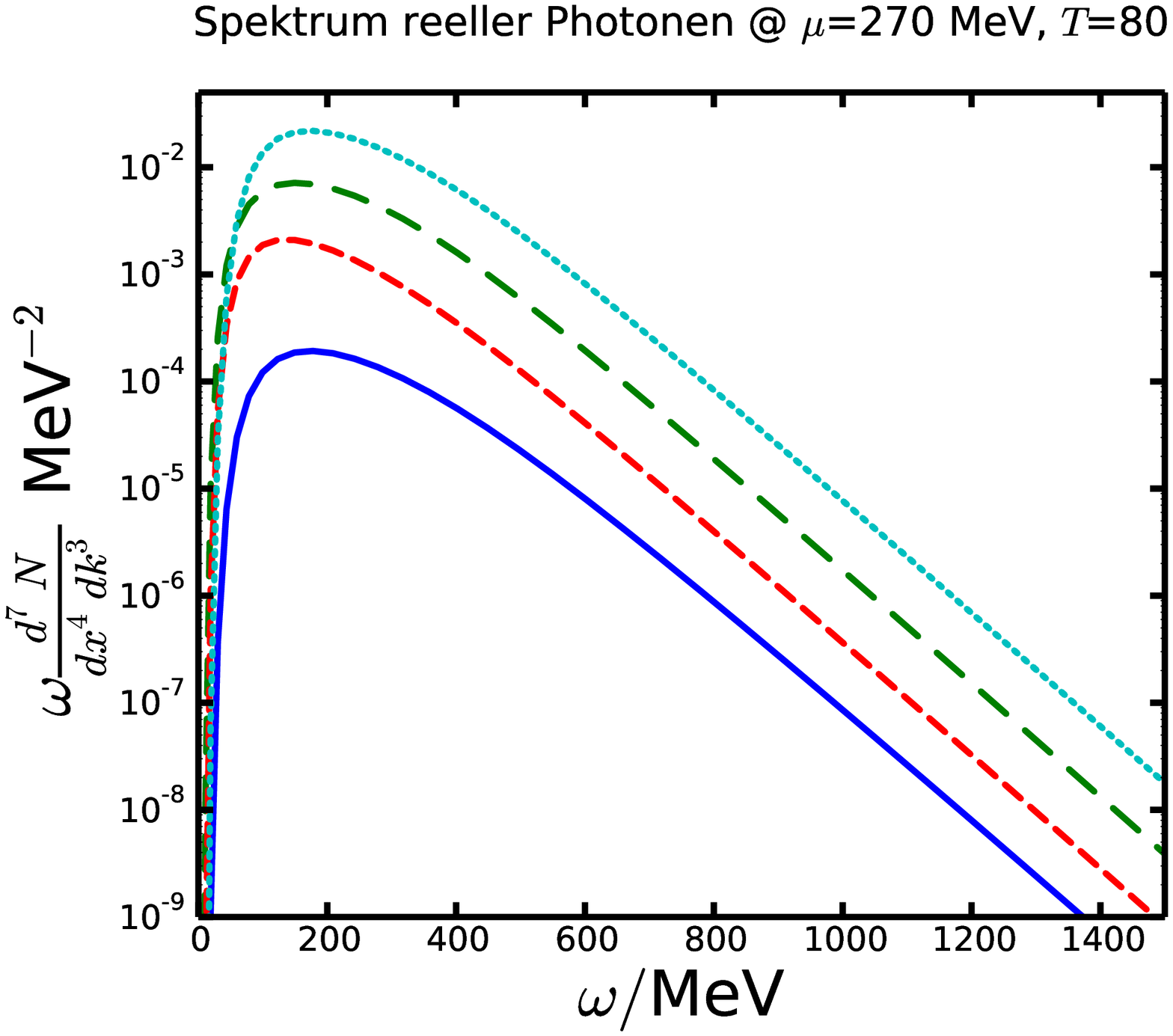}}
             \put(-93, 158){(b)}
             \label{subfig-Spektrum-270-80}}\\
   \subfigure{{\includegraphics[width = 0.49\textwidth,clip=true,trim=4mm 2mm 4mm 16mm]
             {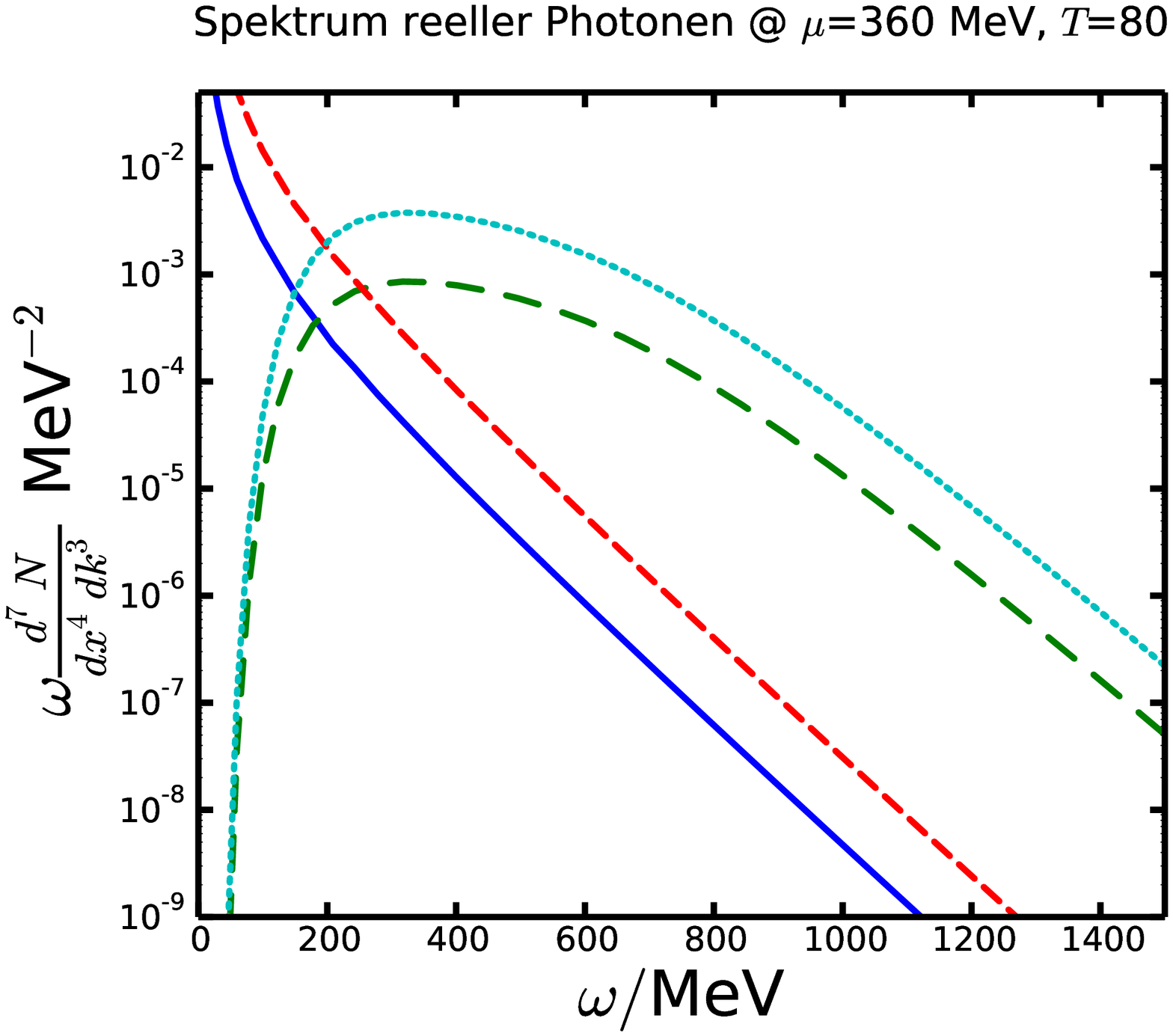}}
             \put(-93, 158){(c)}
             \label{subfig-Spektrum-360-80}}\hfill
   \subfigure{{\includegraphics[width = 0.49\textwidth,clip=true,trim=4mm 2mm 3mm 16mm]
             {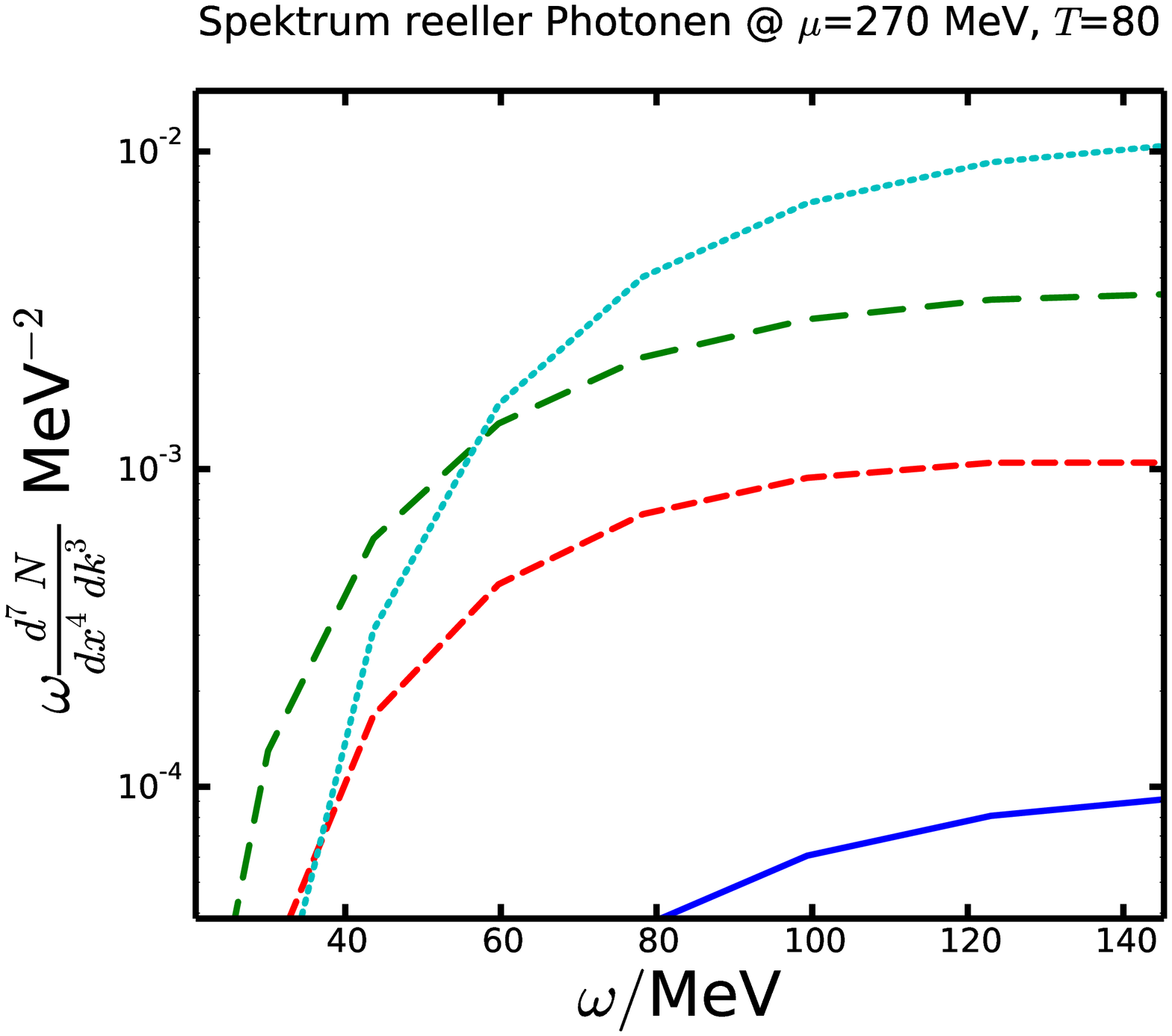}}
             \put(-93, 158){(d)}
             \label{subfig-Spektrum-270-80-zoom}}
   \caption{Photon spectra $\omega\frac{d^7 N}{d^4x d^3 k}$ 
            for the $\LSM$ evaluated at $\mu=0\MeV$ (a), $\mu=270\MeV$ (b) and $\mu=360\MeV$ (c) 
            for $T=80\MeV$; panel (d) is a zoom into (b). 
            Depicted are annihilations into photon and sigma $\psi+\bar\psi\rightarrow \gamma+\sigma$ (solid blue curves)
            and into photon and pions $\psi+\bar\psi\rightarrow \gamma+\pi$ (red short dashed curves) as well as the 
            corresponding Compton processes $\psi+\sigma \rightarrow \psi+\gamma$ (green long dashed curves) and
            $\psi+\pi \rightarrow \psi+\gamma$ (cyan dotted curves). 
            }
   \label{fig-Spektrum}
\end{figure} 
In Fig.~\ref{fig-Spektrum}, the photon spectra originating from four selected processes are plotted for three different
positions in the phase diagram, corresponding to the chirally broken phase (a), the chirally restored phase (c) and the
proximity of the CP (b). 
Comparing the spectra based on \eqref{rate-01}, one notices that in fact some
of the processes develop a maximum at $\approx\omega_p$ and 
vanish at $\omega\rightarrow0$, while others seem to diverge in the IR limit as anticipated above by generalizing
a previous consideration in \cite{Yaresko:2010xe}. 
With the above approximations one can understand why at different positions in the phase diagram the IR behavior of the photon rates
is so different. It is the interplay of the masses of the involved modes and to a less extent the details of the 
interaction process. 
Since for the Compton process the sum of the
incoming masses $m_\psi + m_{\pi,\sigma}$ is larger than the mass of the outgoing quark $m_\psi$, the corresponding contribution
to the photon rate
always has a maximum at finite photon energies. For the annihilation process there is a sensible dependence on the 
position in the phase diagram: In the chirally broken phase, the sigma mass can be larger than twice the quark mass
(\cf the diagonally hatched region in the right panel of Fig.~\ref{fig:massenvgl}). Therefore, there is no maximum 
of the rate at 
non-zero photon energy. Instead, the exponential factor approaches some finite value at $\omega=0$ and other - in the 
former case subleading - effects get dominant. The most prominent effect stems from IR divergencies of the matrix elements
which let the photon rate diverge as $\omega^{-2}$ at $\omega\rightarrow0$ and thus needs to be regularized. 
In a band around the phase transition line and the pseudocritical region 
(horizontally hatched region in the right panel of Fig.~\ref{fig:massenvgl}) 
the meson masses are less than twice the quark mass 
(\eg $m_\psi\sim m_\sigma \sim m_\pi \sim 200\MeV$ at $(T, \mu)\sim (\Tc, \muc)$). Therefore, all processes lead separately to
spectra which vanish at $\omega\rightarrow0$ and have a maximum of similar height and an exponential tail at large $\omega$.
From such a consideration
one can already conclude that a divergency will not appear in a region of the phase diagram where the sigma mass 
is smaller than twice the quark mass.
In the high temperature phase the quarks get light and the mesons heavy (\cf Fig~\ref{fig-massen}).
Therefore, the annihilation rates
have no maxima in the unhatched region of Fig.~\ref{fig:massenvgl} (right panel) and hence 
are not exponentially suppressed at $\omega\rightarrow0$ leading - together with the IR divergency of the
matrix element - to a large photon rate at small $\omega$. 
Thus the pattern of the rate behavior for the subprocesses 
shown in Fig.~\ref{fig-Spektrum} can be explained quite naturally.
Comparing parameter sets with different vacuum sigma masses, the above considerations imply a strong change of the 
low energy photon rates over the phase diagram if we contrast 
the emissivity of the $\LSM$ with low vacuum sigma mass ($m_\sigma^\text{vac} \lesssim 2m_\psi^\text{vac}$) 
to the emissivity at higher vacuum sigma mass ($m_\sigma^\text{vac}\gtrsim 2m_\psi^\text{vac}$). 

\begin{figure}
   \centering
   \subfigure{{\includegraphics[width = 0.49\textwidth,clip=true,trim=5mm 3mm 9mm 14mm]
             {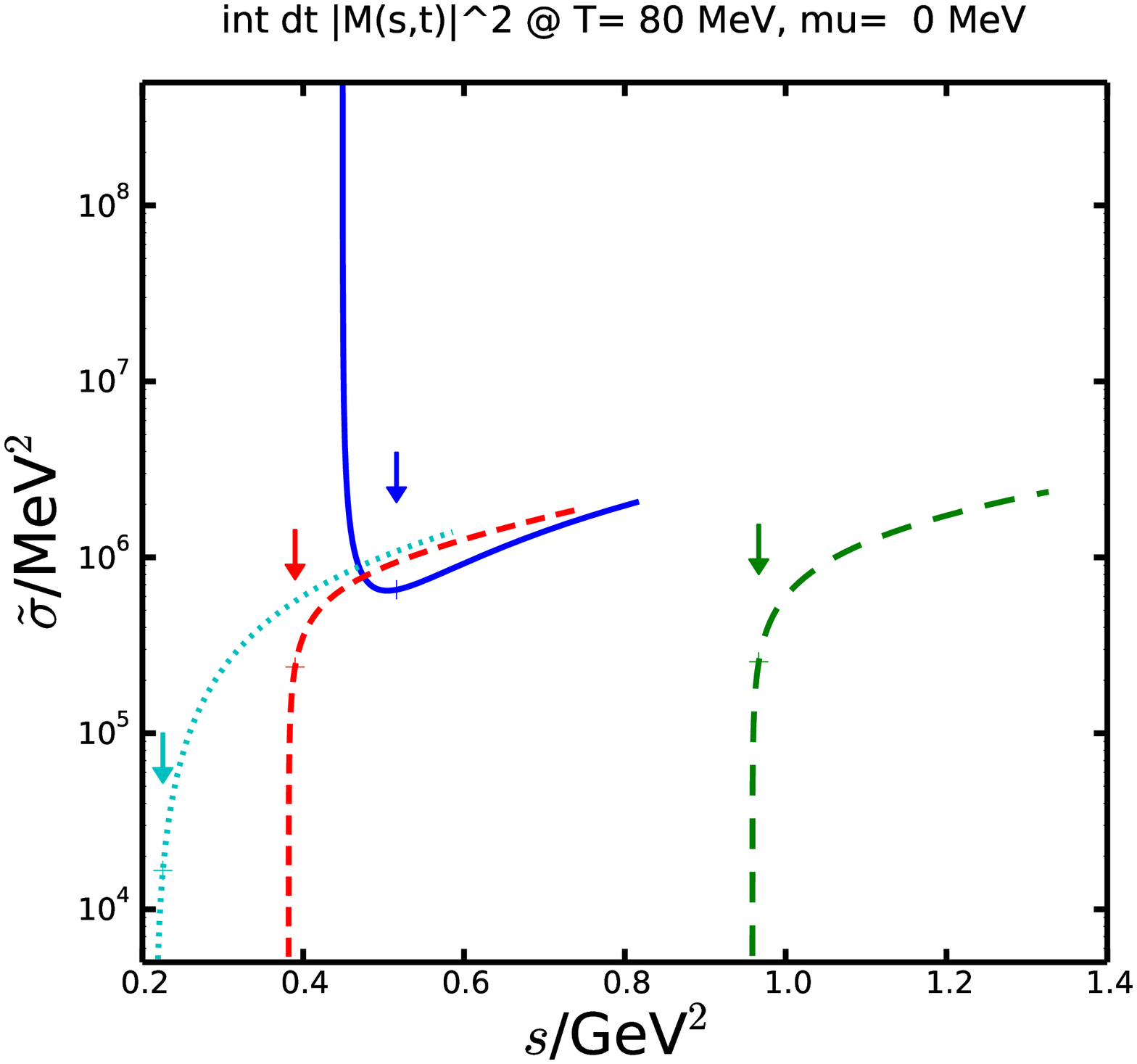}}
             \put(-180, 167){(a)}
             \label{subfig-WQ-000-80}}\hfill
   \subfigure{{\includegraphics[width = 0.49\textwidth,clip=true,trim=6mm 3mm 9mm 14mm]
             {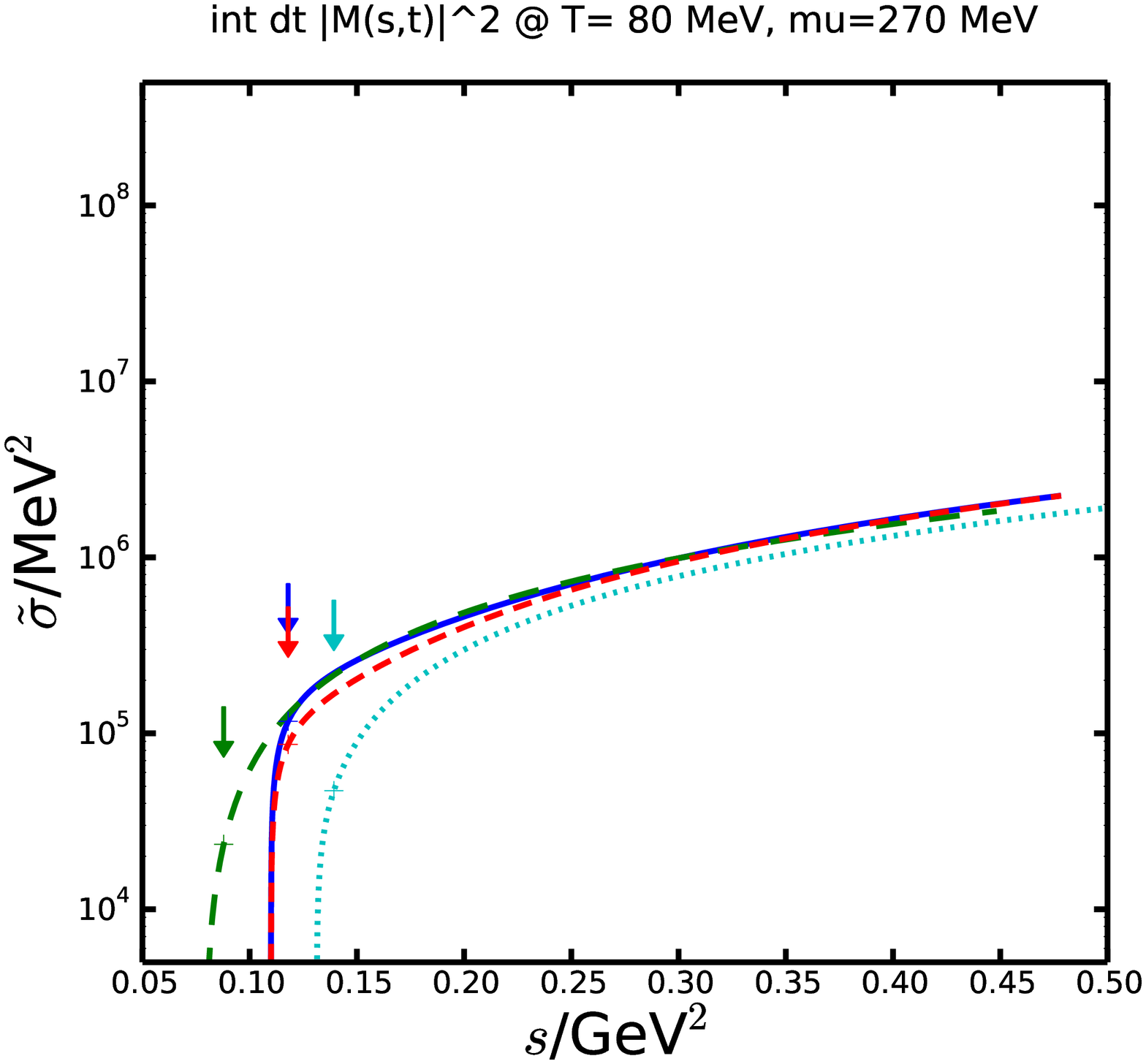}}
             \put(-180, 167){(b)}
             \label{subfig-WQ-270-80}}\\
   \subfigure{{\includegraphics[width = 0.49\textwidth,clip=true,trim=7mm 3mm 9mm 14mm]
             {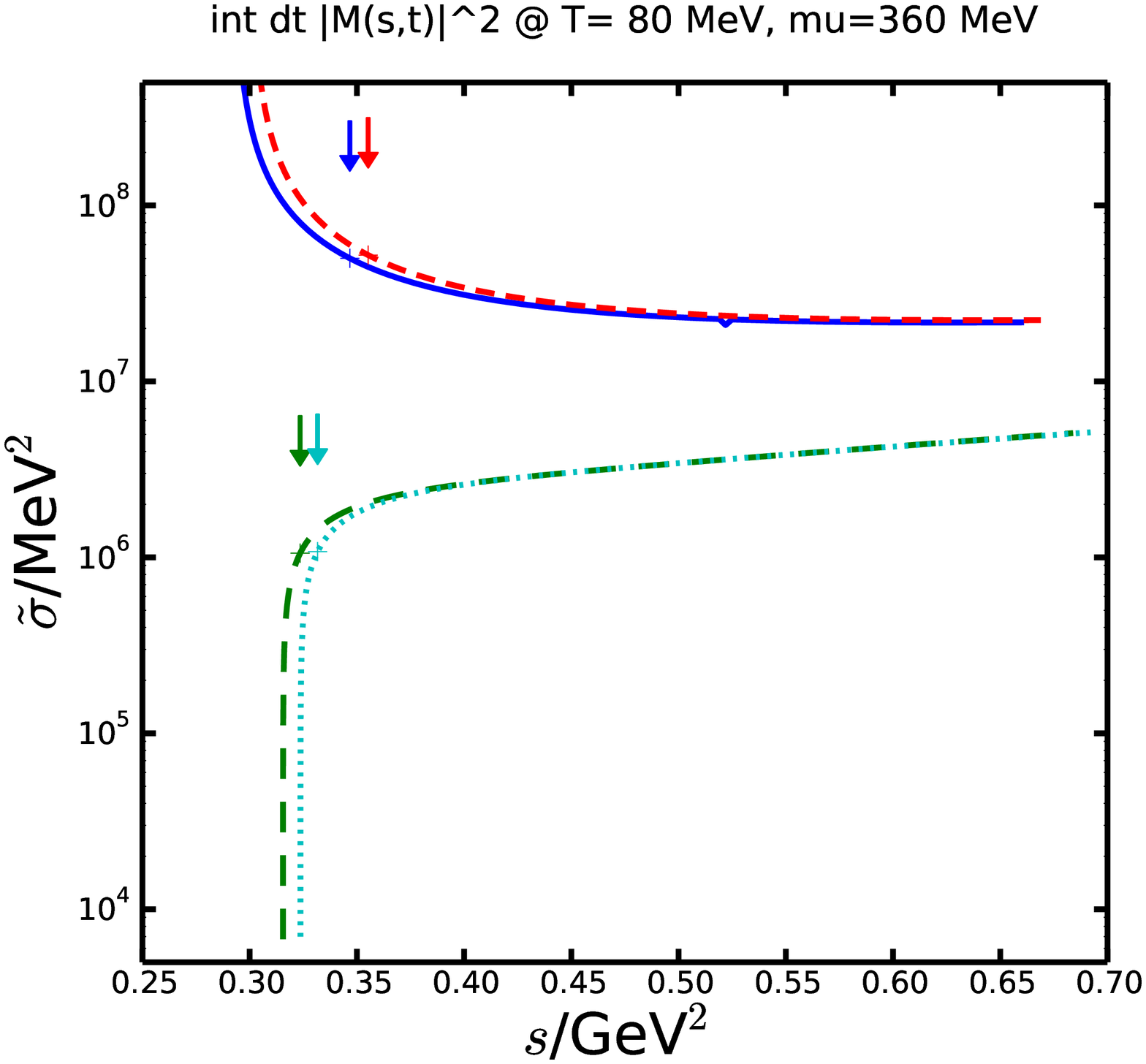}}
             \put(-180, 167){(c)}
             \label{subfig-WQ-360-80}}\hfill
   \begin{minipage}[b]{0.415\textwidth}
      \caption{$\wq$ as a function of $s$ for the reactions exhibited in Fig.~\protect\ref{fig-Spektrum} (same linestyle)
            at the same positions in the phase diagram ($\mu=0, 270\text{ and }360\MeV$ from (a) to (c) at $T=80\MeV$). 
            The arrows depict the maxima of the integrand in \protect\eqref{Boltzmann_01} for $\omega=50\MeV$,
            where the contribution to the respective emissivities are largest.\vspace{27mm}}
      \label{fig-crosssection}
   \end{minipage}
   \hspace{1mm}
\end{figure}

To highlight the role played by the matrix elements we also show in Fig.~\ref{fig-crosssection} the $s$ dependence of 
$\wq$. Again there are two types of functions: If $m_1 + m_2 < m_3$, $\wq(s)$ diverges in the limit 
$s\rightarrow s_0$ like $\sim (s-s_0)^{-1}$, and conversely, if $m_1 + m_2 > m_3$, then $\wq(s)$ behaves like 
$\sqrt{s-s_0}$.
The reason for the divergence in the former case are IR divergencies of the matrix elements, because if 
$s\gtrsim s_0=m_3^2$ the photon energy in the center of mass frame is small and thus the matrix elements are enhanced
because the numerators in the propagators get small.
     
\section{Variation of photon rates over the phase diagram}
The photon rates at fixed photon frequency depend sensitively on the position in the
phase diagram as shown in \cite{Wunderlich:2014cia} for $\omega=10\MeV$. A rate maximum in the critical region for the 
$\sigma$-involving
Compton process was found. We focus here on higher values of $\omega$. From Fig.~\ref{subfig-Spektrum-270-80-zoom}
one notices that up to $\omega\sim50\MeV$ the $\sigma$-involving Compton process (for which the available phase space
around the critical point is enlarged, because of the small sigma mass) is the dominant channel and hence one can 
hope that a signal characteristic for the critical point can be obtained.
\begin{figure}
   \centering
   \subfigure{{\includegraphics[width = 0.49\textwidth,clip=true,trim=7mm 2mm 23mm 15mm]
             {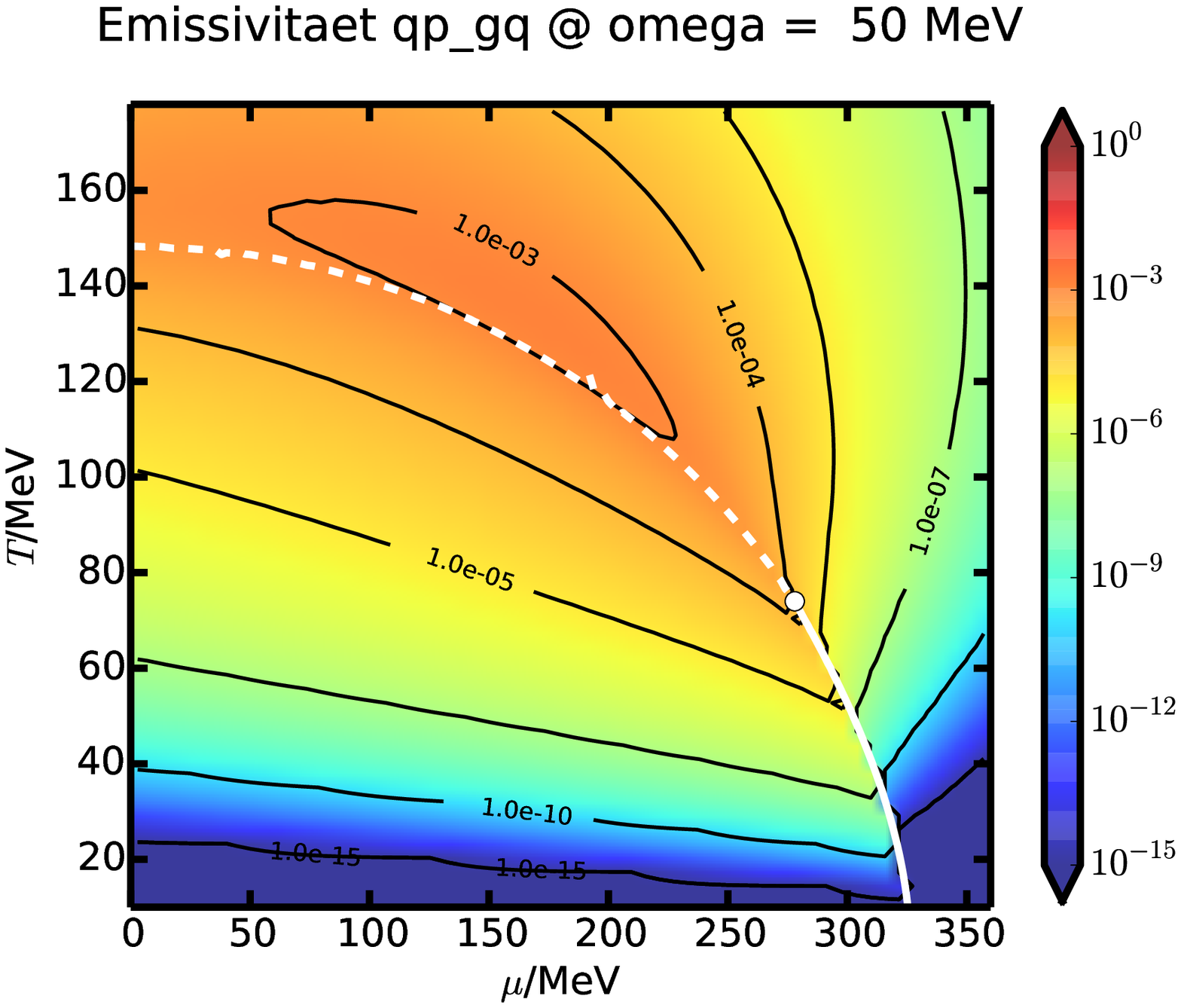}}
             \put(-60, 145){\fcolorbox{black}{white}{(a)}}
             \label{subfig-rate-qp-gq}}\hfill
   \subfigure{{\includegraphics[width = 0.49\textwidth,clip=true,trim=7mm 2mm 23mm 15mm]
             {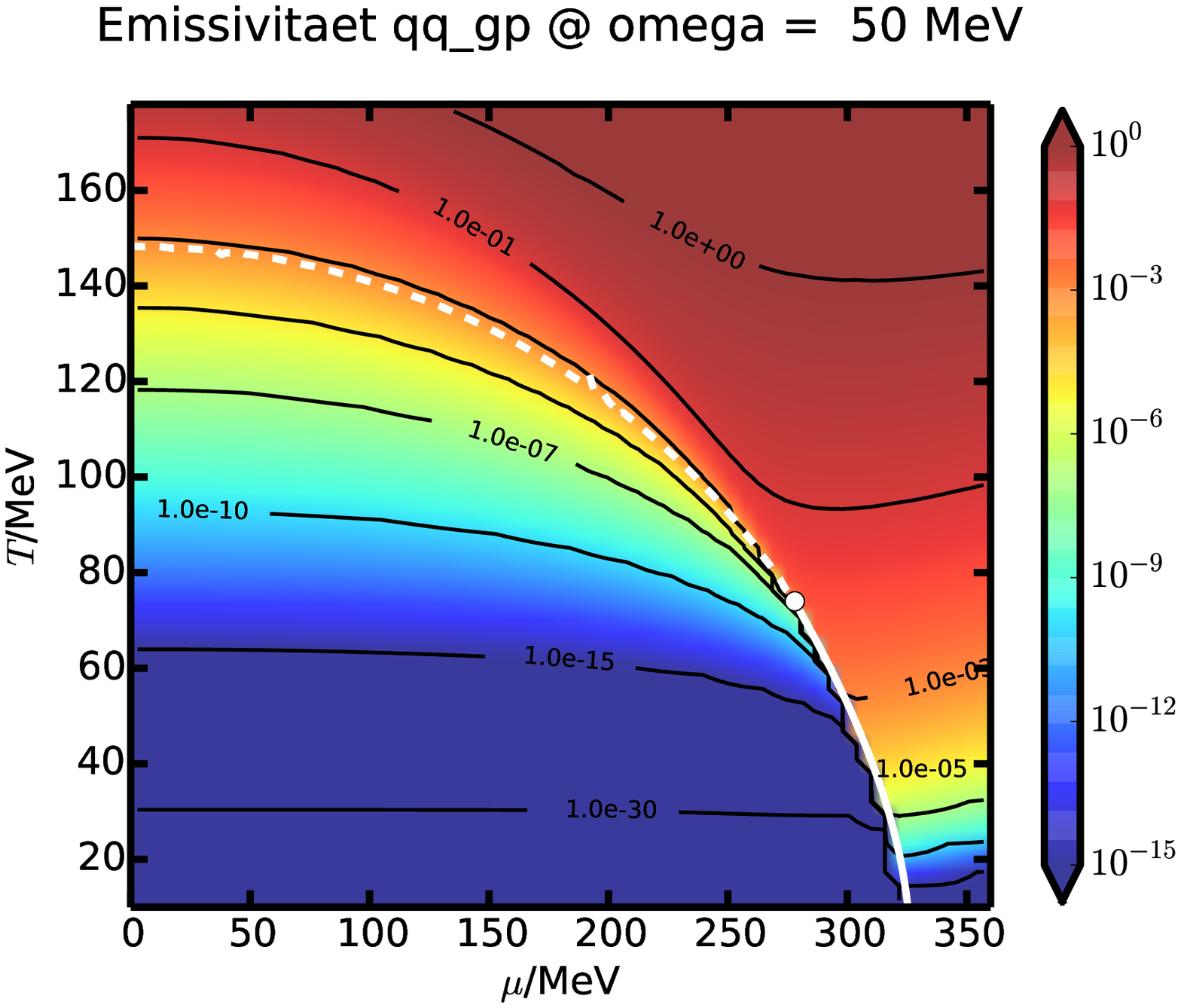}}
             \put(-60, 145){\fcolorbox{black}{white}{(b)}}
             \label{subfig-rate-qq-gp}}\\
   \subfigure{{\includegraphics[width = 0.49\textwidth,clip=true,trim=7mm 2mm 23mm 15mm]
             {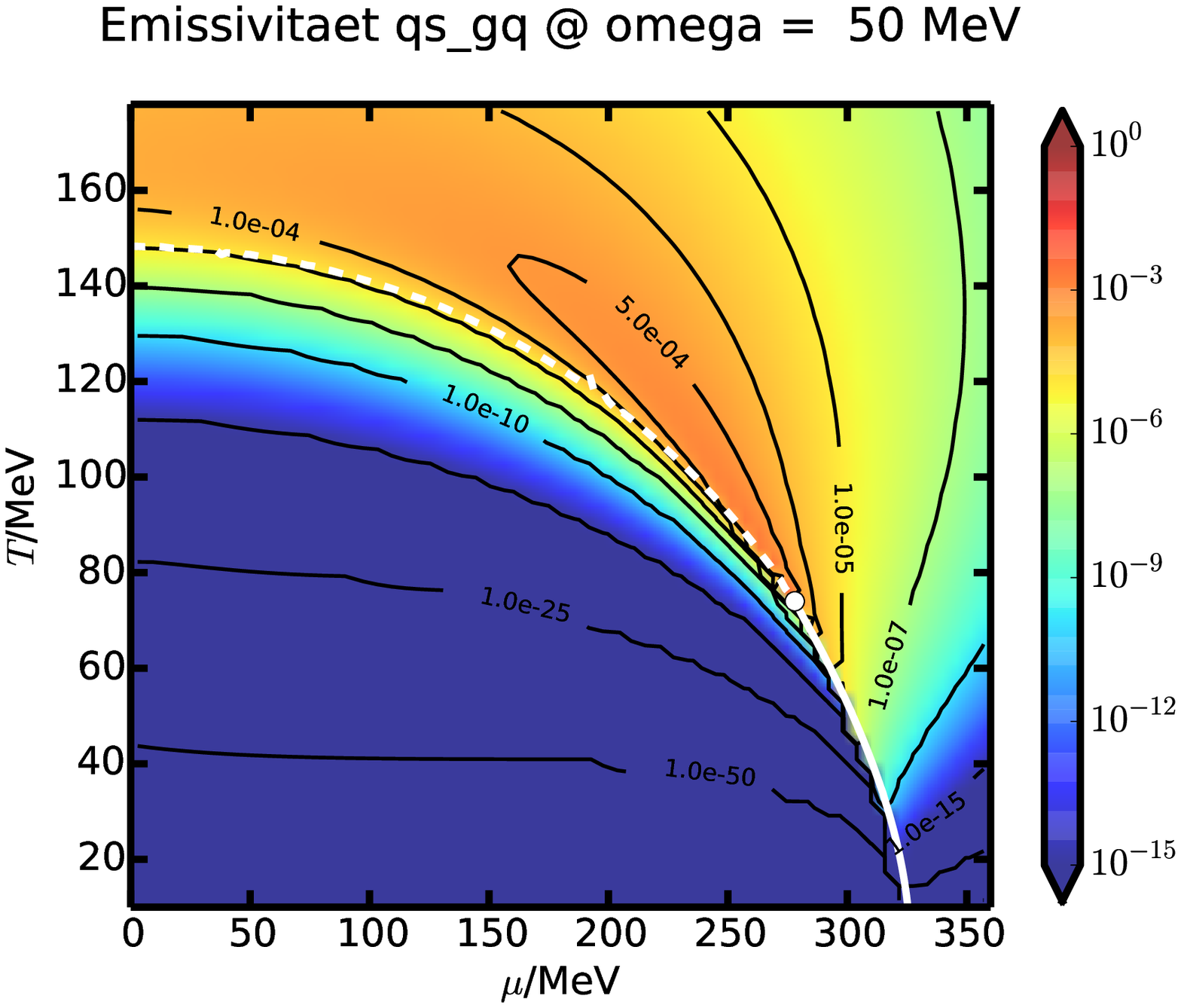}}
             \put(-60, 145){\fcolorbox{black}{white}{(c)}}
             \label{subfig-rate-qs-gq}}\hfill
   \subfigure{{\includegraphics[width = 0.49\textwidth,clip=true,trim=7mm 2mm 23mm 15mm]
             {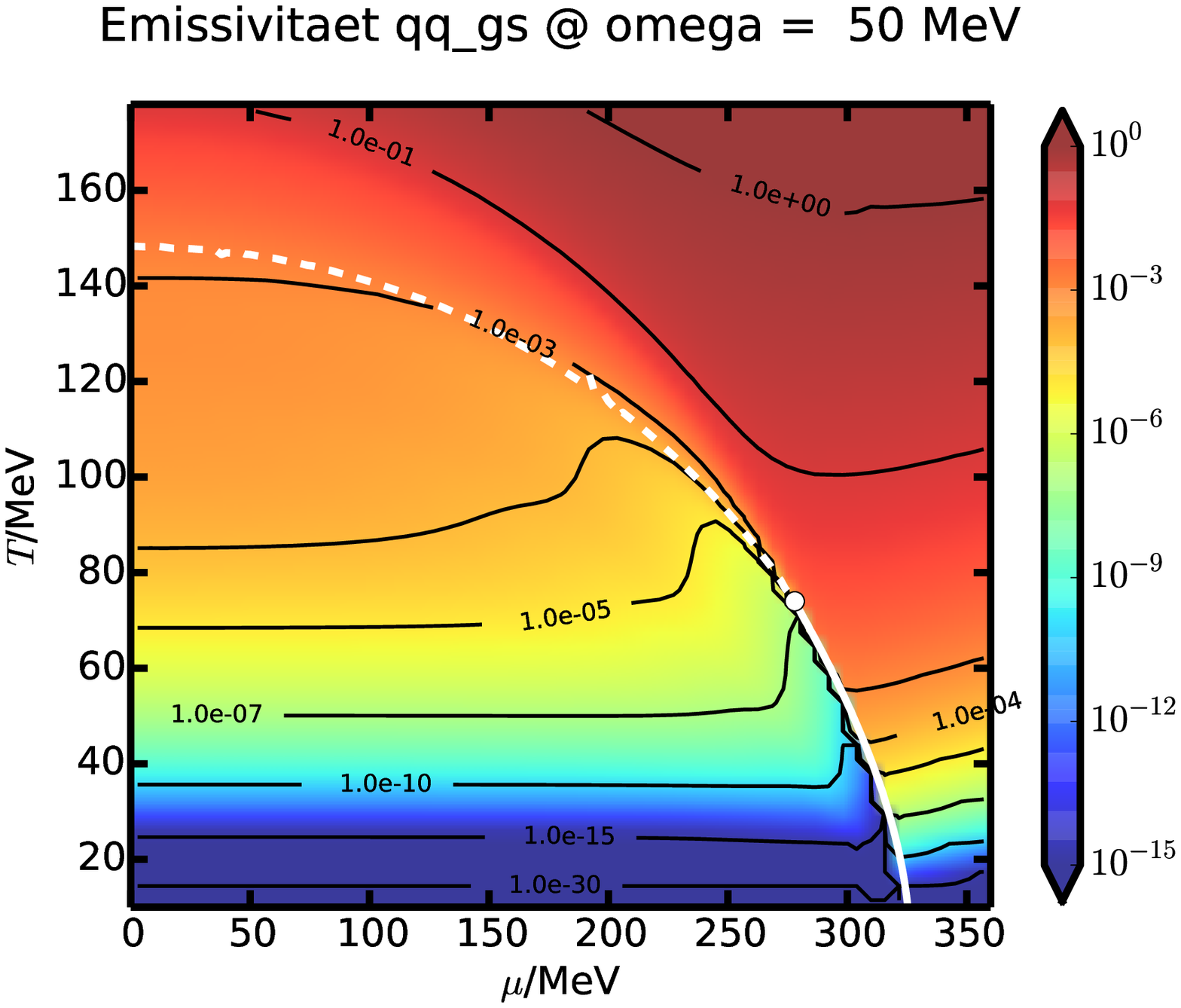}}
             \put(-60, 145){\fcolorbox{black}{white}{(d)}}
             \label{subfig-rate-qq-gs}}
   \caption{
             Contour plots of photon rates $\omega \frac{d^7 N}{d^4x d^3k}$ in units of $\MeV^2$ for $\omega = 50\MeV$
             over the phase diagram for the reactions
             (a): $\psi+\pi\rightarrow\gamma+\psi$, (b): $\psi+\bar\psi\rightarrow\gamma+\pi$, 
             (c): $\psi+\sigma\rightarrow\gamma+\psi$, (d): $\psi+\bar\psi\rightarrow\gamma+\sigma$.
             The phase structure is as in Fig.~\protect\ref{fig:massenvgl} (right panel).
             }
   \label{fig-rates}
\end{figure}
For this frequency, the rates are depicted in Fig.~\ref{fig-rates}. Several 
features are apparent. In the high temperature phase the Compton rates (a) and (c) are suppressed by many orders 
of magnitude relative to the corresponding annihilation processes (b) and (d). While the $\pi$-involving
Compton rate is largest in the crossover region, the $\sigma$-involving Compton rate exhibits a global maximum in the 
critical region. In the low-temperature phase, the annihilation into a $\sigma$ meson and a photon is the strongest 
contributing process and shows the remnant of the structure seen in figure~2 in \cite{Wunderlich:2014cia}.
All of these observations can be explained with the above reasoning. The suppression of the Compton processes is explained
by the different approximation schemes necessary for the two types of processes. Since $m_1+m_2$ is for the Compton process
always larger than $m_3=m_1$, and $\omega=50\MeV$ is far below the peak frequency in the high temperature phase
the exponential suppression factor $\exp\{-F(s_0)/T\}$ is many orders of magnitude smaller than the one for the 
annihilation case $\exp\{-(m_3+\omega)/T\}$.
For $\omega=\Ord{1-2\GeV}$ our results show no special features at $\Tpc$ for small $\mu$.
\section{Summary}
Employing the linear sigma model ($\LSM$) we investigate whether the soft-photon emission rates can reflect the 
conjectured phase structure of QCD. The $\LSM$ is chosen as a simple approach which exhibits a critical point (CP) at 
non-zero chemical potential. Relying on the $\LSM$ field content, which is very schematic and mirrors only in a
limited manner the proper degrees of freedom of QCD, we find, however, that the changes of the quasi-particle 
excitations masses within the phase diagram give rise to significant changes of the emission rates in selected channels.
In particular, the spectral shapes depend strongly on the effective masses of excitation modes involved.
This lets us hope that more advanced considerations can dig out peculiarities of the total emission rate,
\eg caused by the softening of the $\sigma$-type mode near the CP.

The work is supported by BMBF grant 05P12CRGH1.

\bibliography{CPOD_2014}

\end{document}